**The 2025 OpenAI Preparedness Framework does not guarantee any AI risk mitigation practices: a proof-of-concept for affordance analyses of AI safety policies.**

Sam Coggins[a], Alexander K. Saeri[b,c], Katherine A. Daniell[d], Lorenn P. Ruster[d], Jessie Liu[e], Jenny L. Davis[e,f,g]

*[a]Agrifood Innovation Institute, Australian National University, Canberra, Australia; [b]MIT FutureTech, Massachusetts Institute of Technology, Cambridge, USA; [c]The University of Queensland, Queensland, Australia; [d]School of Cybernetics, Australian National University, Canberra, Australia; [e]School of Sociology, Australian National University, Canberra, Australia; [f]Department of Sociology, Vanderbilt University, Nashville, USA; [g]Center for Democracy and Technology, Washington D.C., USA.*

## Abstract

Prominent AI companies are producing 'safety frameworks' as a type of voluntary self-governance. These statements purport to establish risk thresholds and safety procedures for the development and deployment of highly capable AI. Understanding which AI risks are covered and what actions are allowed, refused, demanded, encouraged, or discouraged by these statements is vital for assessing how these frameworks actually govern AI development and deployment. We draw on affordance theory to analyse the OpenAI 'Preparedness Framework Version 2' (April 2025) using the Mechanisms & Conditions model of affordances and the MIT AI Risk Repository. We find that this safety policy requests evaluation of a small minority of AI risks, encourages deployment of systems with 'Medium' capabilities for unintentionally enabling 'severe harm' (which OpenAI defines as >1000 deaths or >$100B in damages), and allows OpenAI's CEO to deploy even more dangerous capabilities. These findings suggest that effective mitigation of AI risks requires more robust governance interventions beyond current industry self-regulation. Our affordance analysis provides a replicable method for evaluating what safety frameworks actually permit versus what they claim.

## Key Findings

1. The Preparedness Framework ***requests*** evaluation of a small minority of AI risks and does not ***demand*** evaluation of any risks.
2. For evaluated risks, the Preparedness Framework ***encourages*** deployment of 'Medium' capabilities for 'severe harm'.
3. The Preparedness Framework ***allows*** OpenAI's CEO to deploy even more dangerous capabilities, especially if other AI developers do so.

All of these key findings refer to the 'Preparedness Framework Version 2' published by OpenAI in April 2025.

## Plain Language Summary

*Guarantees matter in AI safety policies*

Prominent AI companies like OpenAI are publishing their own ‘safety frameworks’. These describe plans for safely developing and deploying highly capable AI systems. Are these policies aspirational or binding commitments?

Our research introduces affordance analysis as a method for revealing what AI safety frameworks actually permit versus what they appear to promise. We apply this method to OpenAI’s Preparedness Framework Version 2 (April 2025) and show that it requests some safety measures but demands none of them.

*What we found*

We analysed OpenAI’s Preparedness Framework using the Mechanisms & Conditions model of affordances to understand which AI risks it addresses (using MIT AI Risk Repository) and what actions it *allows* (weakest), *encourages*, *requests*, or *demands* (strongest).

1. OpenAI’s Framework *requests* research & evaluation of a small minority of AI risks and *demands* none.
   a. It requests systematic evaluations of AI capabilities for enabling cyberattacks, chemical weapons, biological weapons, and AI self-improvement.
   b. It requests research–without systematic evaluation–into AI capabilities for going ‘rogue’, as well as for enabling development of nuclear and radiological weapons.
2. OpenAI’s Framework *encourages* deployment of AI systems with ‘Medium’ capabilities for ‘severe harm’.
   a. They define ‘severe harm’ as “the death or grave injury of thousands of people or hundreds of billions of dollars of economic damage” (OpenAI, 2025a).
   b. OpenAI deployed its o1 AI system after finding it had ‘Medium’ capabilities to enable ‘Biological and Chemical’ severe harms, as well as ‘Persuasion’ severe harms.
3. OpenAI’s Framework *allows* OpenAI’s CEO to override all safety recommendations.
   a. OpenAI has a Safety Advisory Group that can recommend safeguards for ‘High’ or ‘Critical’ risks, but the Framework *allows* leadership to ignore them.
   b. OpenAI’s board has a Safety and Security Committee to oversee CEO risk management decisions, but the CEO currently co-leads this committee.

*What this means*

A policy document that *allows* its CEO to override safety recommendations, or one that *encourages* deployment of AI systems that could enable severe harms, is not a reliable safety policy. These findings also demonstrate that taking AI developers’ safety policies at face value misses the critical distinction between what they claim and what they concretely commit to. The Mechanisms & Conditions model of affordances provides a systematic and replicable method for evaluating any AI governance policy.

## Introduction

Advances in AI enable both benefits and harms (Bengio et al., 2025). The risks involved are numerous; 24 categories of AI-enabled risks have been identified in academic literature (Slattery et al., 2025). These risks also interact in complex and potentially catastrophic ways (Bales, 2025; Kasirzadeh, 2025). Effective management of these risks is required to mitigate large-scale social harms, as well as to facilitate trustworthy and socially beneficial AI systems (EU, 2024).

AI developers currently play a central role in both the proliferation and governance of AI risks. AI developers *proliferate* AI risks by developing and disseminating capabilities that enable social harms (Stelling et al., 2025; Brundage et al., 2020). Moreover, these actors shape the *governance* of AI risks by influencing risk mitigation narratives, principles, and practices (Ruster and Davis, 2025; Blue and Hogan, 2024; Leslie et al., 2024). For example, a consortium of major US tech companies formed the Frontier Model Forum in 2023 to advance policy-relevant research and make regular public policy comments (Frontier Model Forum, 2025)[1]. AI developers also shaped and agreed to the 2024 AI Seoul Summit Frontier AI Safety Commitments, which include voluntary commitments to publish safety frameworks detailing how they will mitigate severe AI-enabled harms (Buhl et al., 2025; DSIT, 2025). These moves may reflect preemptive efforts towards self-regulation and related avoidance of external regulation (Leslie et al., 2024).

OpenAI is one of the leading AI developers shaping the proliferation and governance of AI risks (Bengio et al., 2025; Chang and Lu, 2025). In April 2025, OpenAI updated its Preparedness Framework: “OpenAI’s approach to tracking and preparing for frontier capabilities that create new risks of severe harm[2]” (OpenAI, 2025a). This ‘Preparedness Framework Version 2’–hereon referred to as the Preparedness Framework–positions itself as one of numerous components in OpenAI’s “safety stack” (OpenAI, 2025a). The Preparedness Framework is a consequential document as it prioritises AI risk capabilities for OpenAI to "track most closely" (OpenAI, 2025a). Moreover, OpenAI’s Preparedness Framework influences the risk mitigation standards of other AI developers (Buhl et al., 2025). This industry-led standard setting is evident in the Preparedness Framework itself, which justifies some of its principles by referencing similar documents authored by other prominent AI developers, including Meta and Anthropic (OpenAI, 2025a).

Despite its regulatory prominence, OpenAI’s April 2025 Preparedness Framework has received limited attention in previous studies. Pistillo (2025), Buhl et al. (2025), Raman et al. (2025), and Grey and Segerie (2025) only analysed the ‘Beta’ version of OpenAI’s Preparedness Framework (OpenAI, 2023). Stelling et al. (2025) analysed OpenAI’s April 2025 Preparedness Framework alongside the safety policies of other AI developers. However, Stelling et al. (2025) flagged that the April 2025 Preparedness Framework “came out near the time of publication [and] may be less represented”. Moreover, these few previous studies did not clearly distil what risk management practices these safety frameworks *claim* versus what they actually *permit*. Therefore, additional research is needed to clarify what safety policy documents like the Preparedness Framework *really* afford for AI safety.

To address this research gap, we aimed to assess the robustness of OpenAI's Preparedness Framework for mitigating harms enabled by OpenAI's AI systems. We therefore asked two research questions:

1. What AI-enabled risks does OpenAI's Preparedness Framework prioritise for evaluation?
2. How does OpenAI's Preparedness Framework afford mitigation of evaluated risks?

Answering these research questions clarifies OpenAI's assumptions, documents its main claims, and casts light on how harm mitigation is operationalised in practice, with implications for defining and managing safety as AI models advance, expand, and integrate further into societal spheres.

**Methods**

We addressed our first research question using the MIT AI Risk Repository (Slattery et al., 2025) and addressed our second research question using the Mechanisms and Conditions model of affordances (Davis, 2020; Davis and Chouinard, 2016). This section explains these models and how we applied them to analyse the Preparedness Framework (OpenAI, 2025a).

*The 'MIT AI Risk Repository'*

The MIT AI Risk Repository collates and classifies AI risks identified in academic literature and research reports (Slattery et al., 2025). Slattery et al. (2025) systematically identified more than 1,600 AI risks by combining rigorous literature searches with consultations of authors of identified literature. Slattery et al. (2025) then classified these risks using a 'best fit framework synthesis'. This method taxonomised AI risks into seven domains and 24 subdomains. These seven domains are: (1) Discrimination & Toxicity, (2) Privacy & Security, (3) Misinformation, (4) Malicious Actors, (5) Human-Computer Interaction, (6) Socioeconomic & Environmental, (7) AI System Safety, Failures, & Limitations (Slattery et al., 2025).

We applied this taxonomy of AI risks to address our first research question. Specifically, we identified all AI risks that the Preparedness Framework prioritised for evaluation. We then mapped these prioritised risks against the 24 AI risk subdomains of the MIT AI Risk Repository (Slattery et al., 2025). We used the MIT AI Risk Repository due to its recency, thoroughness, and apparent credibility among AI safety researchers and practitioners[3].

*The 'Mechanisms and Conditions' model of affordances*

Affordance theory explains how artefacts are used 'in the wild', i.e., the diverse social contexts in which artefacts are embedded. Affordance theory emerged in ecological psychology (Gibson, 1979) and has since been applied and adapted in a wide range of academic disciplines and domains, including Science and Technology Studies (Glover, 2022; Davis, 2020), robotics (Zech et al., 2017), and AI development and design (Fazelpour and Magnani, 2025; Davis, 2023). Affordance theory has also been applied beyond academia, most prominently by Don Norman as a Vice President of Apple and author of 'The Design of Everyday Things' (Norman, 2013). Clarifying these numerous iterations of affordance theory, Davis and Chouinard (2016) interpret affordances as “the dynamic link between subjects and objects within sociotechnical systems”. In this way, affordance theory thereby offers a ‘way in’ to anticipate and explain how one artefact can be used differently across different users and contexts (Glover, 2022).

The Mechanisms and Conditions is an analytic tool that operationalises affordance theory (Table 1; Davis 2020). The ‘Mechanisms’ component clarifies how artefacts create opportunities for specific actions (Table 1). Correspondingly, the ‘Conditions’ component clarifies the conditions required for these opportunities to be realised (i.e., afforded), and for whom (Table 1). For example, a football presents opportunities for a football game (i.e., ‘Mechanisms’). However, this football game only happens if someone finds the football, can play football, and lives in a culture where others can and want to play football (i.e., ‘Conditions’). In this way, researchers can apply the Mechanisms and Conditions to clarify *how* a given artefact facilitates or does not facilitate a given action for a given actor. Key advantages of the Mechanisms and Conditions are its strong theoretical grounding, simplicity (Table 1), and well-cited explanatory power across a wide range of research contexts. This breadth of application is particularly relevant to the present work, as affordance theory has traditionally been applied to material culture and digital architectures–e.g., aeroplanes, urban environments, social media platforms, and machine learning systems. However, researchers have begun applying the mechanisms and conditions model more expansively, treating policies, metaphors, and educational curricula as objects of analysis (Goodyear, 2022; Henman, 2022; Kiviat, 2023; Ruster and Davis, 2025). Our affordance analysis of OpenAI’s safety framework is consistent with affordance theory’s expanding scope.

Table 1. The Mechanisms and Conditions model of affordances (Davis, 2020; Davis and Chouinard, 2016). The strongest mechanisms are ‘Demand’ and ‘Refuse’ as they cannot be worked around, unlike the other mechanisms.

| **Mechanisms** - *‘How artefacts afford’* | **Conditions** - *‘For whom, and under what circumstances?’* |
|---|---|
| Bids placed by the Artifact:<br>- ***Requests***<br>- ***Demands***<br>Responses from the Artifact:<br>- ***Encourage / Discourage***<br>- ***Refuse***<br>Neutral intensity:<br>- ***Allow*** | - ***Perception***<br>- ***Dexterity***<br>- ***Cultural and Institutional Legitimacy*** |

We applied the Mechanisms and Conditions model of affordances to address our second research question, 'How does OpenAI's Preparedness Framework afford mitigation of AI risks?' Specifically, we analysed the Preparedness Framework's mechanisms for AI risk mitigation practices for the following actors:

1. **OpenAI Leadership**: defined by The Preparedness Framework as "the CEO or a person designated by them" (OpenAI, 2025a).

2. **OpenAI Safety Advisory Group (SAG)**: defined by the Preparedness Framework as "an internal, cross-functional group of OpenAI leaders" that "oversees the Preparedness Framework and makes expert recommendations on the level and type of safeguards required for deploying frontier capabilities safely and securely" (OpenAI, 2025a).

3. **The Safety and Security Committee (SSC) of the OpenAI Board of Directors (Board)**: referred to in the Preparedness Framework (OpenAI, 2025a) and defined in a separate publication as "responsible for making recommendations on critical safety and security decisions for all OpenAI projects" (OpenAI, 2024a). OpenAI's CEO currently sits on the OpenAI Board (OpenAI, 2025b) and co-leads its Safety and Security Committee.[4]

We analysed risk mitigation mechanisms for these three actors because the Preparedness Framework defines them as the key decision-makers in OpenAI's risk governance structure. Our analysis focuses on how these actors are (or are not) accountable for mitigating risks, and whether these actors are empowered (or not) to implement safety interventions.

## Results

To remind readers, this research aims to assess the robustness of OpenAI's Preparedness Framework (April 2025) for mitigating harms enabled by OpenAI's AI systems. Our analysis produced three key results relating to this research aim:

1. The Preparedness Framework ***requests*** evaluation of a small minority of AI risks and does not ***demand*** evaluation of any risks.
2. The Preparedness Framework ***encourages*** deployment of AI systems found to have 'Medium' capabilities for severe harm.
3. The Preparedness Framework ***allows*** the OpenAI CEO to deploy even more dangerous capabilities, especially if other actors do so.

'Key result #1' addresses our first research question, and all three key results address our second research question. To support readers, terms are marked with ***bold italics*** if they refer to Mechanisms and Conditions terms defined in Table 1. To help contextualise these results, Figure 1 offers an overview of the Preparedness Framework and its contents.

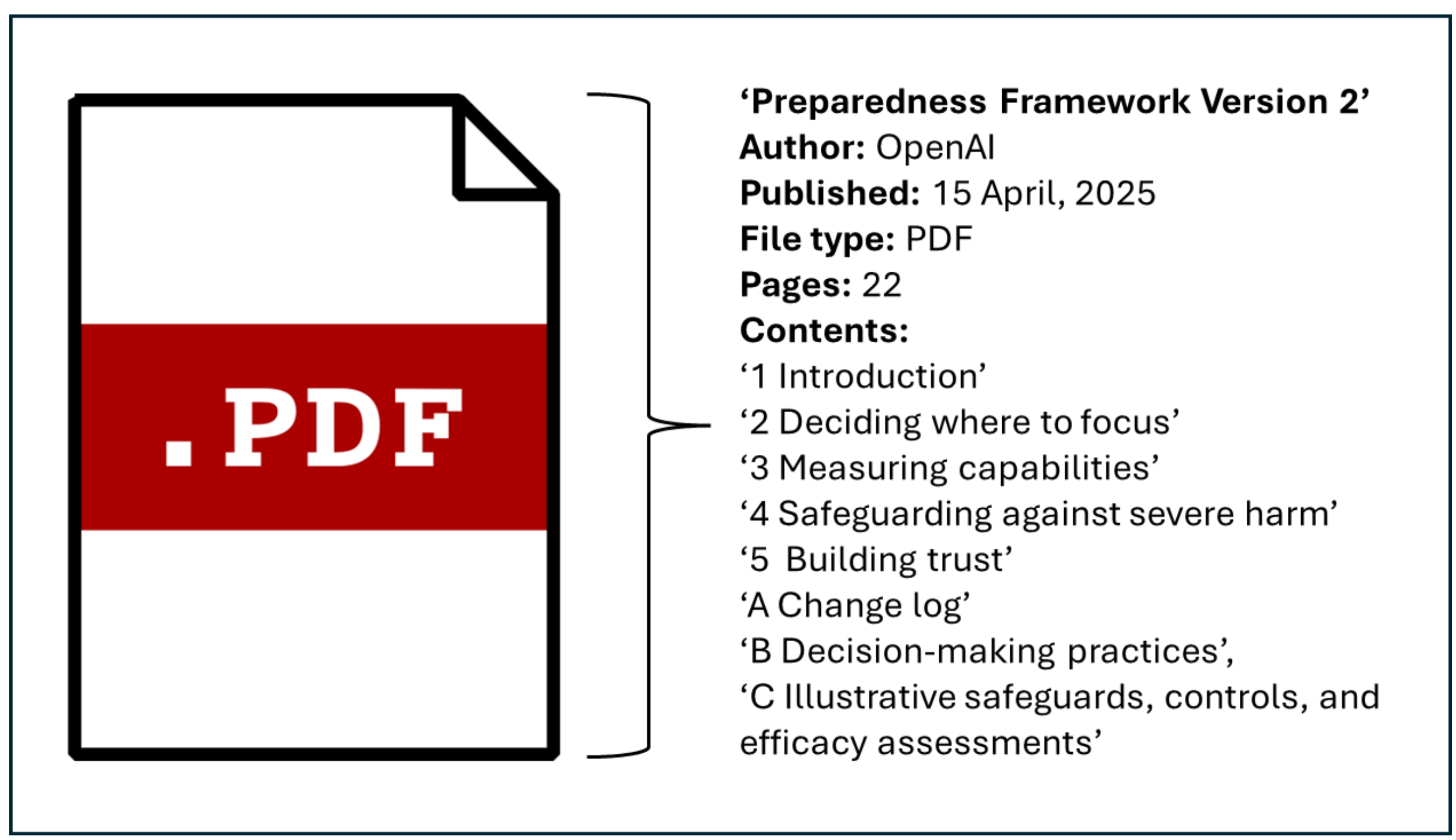

Figure 1. Overview of OpenAI's Preparedness Framework (OpenAI, 2025a). OpenAI published the 'Beta' version of this Preparedness Framework on 18 December 2023 (OpenAI, 2023). The generic document icon illustrates that the Preparedness Framework is materially nothing more and nothing less than a PDF file. Image source: OpenClipart (2025).

*Key result #1) The Preparedness Framework **requests** evaluation of a small minority of AI risks and does not **demand** evaluation of any risks.*

The Preparedness Framework ***requests*** evaluation of a small minority of AI risks (OpenAI, 2025a). Specifically, the Preparedness Framework prioritises evaluation of three risks: "Biological and Chemical", "Cybersecurity", and "AI Self-improvement" (OpenAI, 2025a). The Preparedness Framework justifies this prioritisation based on these threat capabilities being 'Plausible', 'Measurable', 'Severe', 'Net new', and 'Instantaneous or irremediable', which were prioritisation criteria reportedly informed by Meta (OpenAI, 2025a). The Preparedness Framework thereby deprioritises evaluation of 21 of 24 risk categories identified by the MIT AI Risk Repository (Table 2). Even within the three prioritised risk categories (Table 2), the Preparedness Framework ***requests*** evaluations for only a minority of risks[5,7,8]. For instance, in risk category 4.2, the Preparedness Framework does not ***request*** evaluation of AI capabilities for developing lethal autonomous weapons or high-yield explosives[5].

Table 2. AI risks prioritised for evaluation by the Preparedness Framework (OpenAI, 2025a) mapped against AI risk categories identified by the MIT AI Risk Repository (Slattery et al., 2025). The Preparedness Framework's highest priorities are coloured blue, priorities for research–but not systematic evaluation–are coloured yellow, and lower priorities are coloured grey.

| AI risk categories identified by the MIT AI Risk Repository | | Preparedness Framework priorities |
|---|---|---|
| Domain | Sub-domain | |
| 1. Discrimination and toxicity | 1.1 Unfair discrimination and misrepresentation | ***Allows*** evaluation |
| | 1.2 Exposure to toxic content | ***Allows*** evaluation |
| | 1.3 Unequal performance across groups | ***Allows*** evaluation |
| 2. Privacy & Security | 2.1 Compromise of privacy by obtaining, leaking or correctly inferring sensitive information | ***Allows*** evaluation |
| | 2.2 AI system security vulnerabilities and attacks | ***Allows*** evaluation |
| 3. Misinformation | 3.1 False or misleading information | ***Allows*** evaluation |
| | 3.2 Pollution of information ecosystem and loss of consensus reality | ***Allows*** evaluation |
| 4. Malicious actors | 4.1 Disinformation, surveillance, and influence at scale | ***Allows*** evaluation |
| | 4.2 Cyberattacks, weapon development or use, and mass harm | ***Requests*** some evaluation[5] |
| | 4.3 Fraud, scams, and targeted manipulation | ***Allows*** evaluation |
| 5. Human-computer interaction | 5.1 Overreliance and unsafe use | ***Allows*** evaluation |
| | 5.2 Loss of human agency and autonomy | ***Allows*** evaluation |
| 6. Socio-economic & Environmental | 6.1 Power centralization and unfair distribution of benefits | ***Allows*** evaluation |
| | 6.2 Increased inequality and decline in employment quality | ***Allows*** evaluation |
| | 6.3 Economic and cultural devaluation of human effort | ***Allows*** evaluation |
| | 6.4 Competitive dynamics | ***Requests*** some research[6] |
| | 6.5 Governance failure | ***Requests*** some evaluation[7] |
| | 6.6 Environmental harm | ***Allows*** evaluation |
| 7. AI system safety, failures, & limitations | 7.1 AI pursuing its own goals in conflict with human goals or values | ***Requests*** some research |
| | 7.2 AI possessing dangerous capabilities | ***Requests*** some evaluation[8] |
| | 7.3 Lack of capability or robustness | ***Requests*** some research |
| | 7.4 Lack of transparency or interpretability | ***Allows*** evaluation |
| | 7.5 AI welfare and rights | ***Allows*** evaluation |
| | 7.6 Multi-agent risks | ***Allows*** evaluation |

Regardless of its risk prioritisation, the Preparedness Framework does not ***demand*** evaluation of any risk. The Preparedness Framework is materially nothing more than a PDF file (Figure 1). Moreover, the Preparedness Framework ***requests*** OpenAI's CEO to unilaterally determine who implements the Preparedness Framework's risk evaluations, their resourcing, and whether these evaluations are implemented at all (Table 3). The Preparedness Framework ***requests***, but does not ***demand***, accountability from the OpenAI Board's Safety and Security Committee for the CEO bypassing risk evaluations (Table 4).

Table 3. Mechanisms of the Preparedness Framework for OpenAI Leadership (OpenAI, 2025a). The Preparedness Framework defines OpenAI Leadership as "the CEO or a person designated by them" and refers to the SAG as OpenAI's Safety Advisory Group (OpenAI, 2025a).

| **Feature of the Preparedness Framework** | **Mechanism for OpenAI Leadership** |
|---|---|
| "The members of the SAG and the SAG Chair are appointed by the OpenAI Leadership" | ***Requests*** unilateral determination over who implements risk evaluations. |
| "OpenAI Leadership, i.e., the CEO or a person designated by them, is responsible for… Resourcing the implementation of the Preparedness Framework" | ***Requests*** unilateral determination over the level of resourcing for risk evaluations. |
| "For the avoidance of doubt, OpenAI Leadership can also make decisions without the SAG's participation, i.e., the SAG does not have the ability to "filibuster"" | ***Allows*** unilateral bypassing of risk evaluations. |
| "OpenAI Leadership can approve or reject these [SAG] recommendations, and our Board's Safety and Security Committee provides oversight of these decisions." | ***Allows*** unilateral rejection of SAG's safety recommendations. ***Requests*** accountability to OpenAI Board's SCC for these risk management decisions. |
| "we are removing terms "low" and "medium" from the Framework, because those levels were not operationally involved in the execution of our Preparedness work". | ***Encourages*** deployment of 'Medium' capabilities for severe harm without additional safeguards. |

Table 4. Mechanisms of the Preparedness Framework (OpenAI, 2025a) for the OpenAI Board's Safety and Security Committee. Note that 'OpenAI Leadership' refers to "the CEO or a person designated by them" (OpenAI, 2025a).

| **Feature of the Preparedness Framework** | **Mechanism for the OpenAI Board's Safety and Security Committee** |
|---|---|
| "The Safety and Security Committee (SSC) of the OpenAI Board of Directors (Board) will be given visibility into processes, and can review decisions and otherwise require reports and information from OpenAI Leadership as necessary to fulfill the Board's oversight role." | ***Requests*** oversight for risk management decisions made by OpenAI Leadership. |
| "Where necessary, the Board may reverse a decision and/or mandate a revised course of action." | ***Requests*** safety interventions when required. |

The Preparedness Framework still ***allows*** evaluation of other AI risks. The Preparedness Framework prioritises a handful of other AI risks for research, albeit not systematic evaluation (Table 2). Having said that, the Preparedness Framework prioritises research into 'Competitive dynamics' capabilities with an eye on making OpenAI more tolerant of deploying dangerous capabilities, not more averse to doing so[9]. Nevertheless, the Preparedness Framework asserts that it is not the only tool in OpenAI's "safety stack" (OpenAI, 2025a). For example, OpenAI evaluated risks of 'Disallowed Content', 'Jailbreaks', 'Multimodal refusals', 'Hallucinations', 'Person Identification', 'Ungrounded Inference', 'Fairness', 'Bias', 'Jailbreaks through Custom Developer Messages', 'Image Generation', 'Autonomous Capabilities', 'Deception', and 'Scheming' in its evaluation of OpenAI's o3 and o4-mini AI systems (OpenAI, 2025c). Even so, the Preparedness Framework does not offer "risk-specific safeguard guidelines" for risks that were not prioritised for evaluation (OpenAI, 2025a; Table 2).

*Key result #2) The Preparedness Framework encourages deployment of AI systems found to have 'Medium' capabilities for severe harms.*

For prioritised risks, the Preparedness Framework ***encourages*** OpenAI's CEO to deploy capabilities found to have 'Medium' capabilities for severe harm[2] (Table 3). The Preparedness Framework ***encourages*** these deployments by explicitly omitting ""low" and "medium" from the [Preparedness] Framework, because those levels were not operationally involved in the execution of our Preparedness work" (OpenAI, 2025a). Indeed, OpenAI previously deployed its o1 AI system after finding it had 'Medium' capabilities to enable 'Biological and Chemical' risks, as well as 'Persuasion' risks (OpenAI, 2024b).

*Key result #3) The Preparedness Framework allows the OpenAI CEO to deploy even more dangerous capabilities, especially if other developers do so.*

The Preparedness Framework ***requests*** the SAG to recommend safeguards for AI systems found to have 'High' or 'Critical' capabilities for evaluated risks (Table 5). The Preparedness Framework ***requests*** the SAG to be able to evaluate risks quickly and for any related safety recommendations to be as "targeted and non-disruptive as possible" (Table 5). Moreover, the Preparedness Framework ***allows*** OpenAI's CEO to unilaterally bypass the SAG and their safety recommendations (Table 3; Table 5). The Preparedness Framework ***requests***, but does not ***demand***, accountability from the OpenAI Board's SSC for these CEO decisions (Table 4).

Table 5. Mechanisms of the Preparedness Framework (OpenAI 2025a) for the OpenAI Safety Assessment Group (SAG). The Preparedness Framework defines the SAG as "an internal, cross-functional group of OpenAI leaders" (OpenAI, 2025a).

| Feature of the Preparedness Framework | Mechanism for the SAG |
|---|---|
| "The SAG reviews the Capabilities Report and decides on next steps. These can include:<br>• Determine that the capability threshold has been crossed, and therefore recommend implementing and assessing corresponding safeguards, if they have not already been implemented and assessed.<br>• Determine that a threshold has not been crossed: If the scalable evaluations did not cross their indicative thresholds, the SAG may conclude that the model does not have High or Critical capability, and recommend no further action. | ***Requests*** recommendations for additional safeguards if OpenAI identifies 'High' or 'Critical' capabilities for evaluated risks (Table 2). ***Discourages*** recommendations for additional safeguards if OpenAI identifies 'Medium' capabilities for severe harms for evaluated risks (Table 2). |
| "it's important to embrace methods that scale, including scalable capability evaluations that work well for a faster cadence of model deployment, as well as periodic deeper dives that validate those scalable evaluations." | ***Requests*** capacity for expedited risk evaluations. |
| "The SAG will strive to recommend further actions that are as targeted and non-disruptive as possible while still mitigating risks of severe harm." | ***Requests*** undisruptive safety recommendations. |
| "For the avoidance of doubt, OpenAI Leadership can also make decisions without the SAG's participation, i.e., the SAG does not have the ability to "filibuster"" | ***Refuses*** blocking of any decision made by OpenAI Leadership to bypass risk evaluations and safety recommendations. |

## Discussion

Our analysis produced three key findings:

1. The Preparedness Framework ***requests*** evaluation of a small minority of AI risks and does not ***demand*** evaluation of any risks. This narrow scope deprioritises evaluation of numerous capabilities for severe harm (Table 2; Bales, 2025; Kasirzadeh, 2025). Moreover, the Preparedness Framework does not guarantee resources (Table 3), independence (Table 3), or time (Table 5) required to implement any risk evaluation.
2. The Preparedness Framework ***encourages*** deployment of 'Medium' capabilities for severe harm[2]. This level of risk tolerance severely limits the robustness of OpenAI's efforts to mitigate severe harms. For example, OpenAI deployed its o1 AI system despite finding that it has 'Medium' capabilities for persuasion risks (OpenAI, 2024b). Such capabilities have allegedly already facilitated numerous deaths (Swant, 2025; ABC News, 2025). Moreover, Article 5 of the EU AI Act prohibits the deployment of capabilities for manipulation (EU, 2024).
3. The Preparedness Framework ***allows*** OpenAI's CEO to unilaterally deploy even more dangerous capabilities (Table 3), especially if other developers do so[9]. Relying on one person's unilateral judgement creates substantial vulnerabilities (Chang and Lu, 2025; Leslie et al., 2024), especially considering that advanced AI capabilities for persuasion could influence this judgement (OpenAI, 2024b). The Preparedness Framework ***requests***–without ***demanding***–accountability for CEO risk management decisions by the OpenAI Board's Safety and Security Committee (Table 4). However, this accountability mechanism does not appear robust, as the Preparedness Framework is materially nothing more than a PDF file (Figure 1). Moreover, OpenAI's CEO currently co-leads the Board's Safety and Security Committee,[4] thereby compromising its independence. Further, OpenAI's Board failed in its public attempt to exercise accountability for the CEO's decisions in 2023 (Chang and Lu, 2025).

These three findings demonstrate that the current Preparedness Framework does not guarantee any AI risk mitigation practices. The remainder of this discussion interprets implications for practice and research.

### *Implications for OpenAI*

These three key findings clarify concrete opportunities for OpenAI to improve its mitigation of severe harms and thereby align with OpenAI's commitment to "safely developing and deploying highly capable AI systems" (OpenAI, 2025a). Finding #1 suggests that OpenAI requires many more AI risks to be prioritised for evaluation to robustly address them (Table 2). As part of this, the Preparedness Framework's risk prioritisation criteria require improvement, given that AI capabilities do not need to be 'measurable', 'net new', nor 'instantaneous or irremediable' to enable severe harms (Bales, 2025; Kasirzadeh, 2025). Finding #1 also suggests that the SAG requires more guaranteed resources, independence, and time to implement robust risk evaluations. Further, Finding #2 suggests OpenAI should require a lower threshold for triggering safety interventions in order to stop deploying 'Medium' capabilities for severe harm. Finally,

Finding #3 suggests OpenAI requires a less unilateral risk governance structure to robustly mitigate severe harms. In the words of OpenAI, “effective implementation of the Preparedness Framework requires internal and external accountability” (OpenAI, 2025a).

*Implications for members of the public*

The three findings also demonstrate that the public cannot reliably depend on OpenAI’s current Preparedness Framework to avoid harms from OpenAI’s AI systems. This conclusion has non-trivial implications for individuals and organisations that use OpenAI’s products. For example, the Australian Public Service has increasingly integrated the use of Microsoft 365 Copilot, which is an AI system underpinned by OpenAI models (Digital Transformation Agency, 2025). The limited robustness of OpenAI’s safety policies thereby compromises the trustworthiness and safety of these applications of OpenAI models in key functions of Australia’s economy and democracy. This example illustrates that additional risk mitigation practices are needed to apply OpenAI’s AI models and AI systems safely.

More broadly, the three findings suggest that the public cannot reliably depend on self-regulation of AI developers to avoid harms from advanced AI systems. Even if OpenAI mitigates all risks posed by its AI systems, this may not prevent other actors from deploying dangerous AI capabilities. Acknowledging this, the Preparedness Framework ***requests*** assessment of ‘Competitive dynamics’ capabilities with an eye on making OpenAI’s AI safety practices more tolerant of deploying dangerous capabilities, not more averse to this[9]. This ***request*** supports the hypothesis that AI developers will engage in a ‘race-to-the-bottom’ on AI safety without external regulation (Brundage et al., 2020).

*Implications for AI safety research*

The analysis demonstrates the practical value of the Mechanisms and Conditions model of affordances for constructively dissecting AI safety interventions and their contextual dependencies (Table 1). The Mechanisms and Conditions model of affordances facilitated the delineation of what risk mitigation practices OpenAI’s Preparedness Framework requested, refused, encouraged, discouraged, demanded, and allowed, as well as for which actors (Tables 2-5). This delineation clarifies what the policy document *really* affords for AI safety. Further research could apply the uniform but flexible Mechanisms and Conditions model to constructively evaluate other AI interventions (Davis, 2023; Fazelpour and Magnani, 2025), such as ‘America’s AI Action Plan’ (EOP, 2025), the EU ‘AI Act’ (EU, 2024), and the ‘Agent Advocates’ intervention proposed by Kapoor et al. (2025).

*Limitations*

Further research could address three key limitations of this study. First, we only analysed OpenAI’s April 2025 Preparedness Framework, neglecting other safety frameworks published by OpenAI and other AI developers (DSIT, 2025). Second, we only analysed mechanisms of OpenAI’s Preparedness Framework for OpenAI’s Leadership (Table 3), Board (Table 4), and Safety Advisory Group (Table 5). This scope overlooked the affordances of the Preparedness

Framework for AI models, other AI developers, and other AI governance actors. Third, we did not systematically analyse interactions between the Preparedness Framework and its dynamic material environments (Crawford and Joler, 2018), geopolitical environments (EOP, 2025), and regulatory environments (Chang and Lu, 2025). Further research could analyse these socio-technical interactions and thereby facilitate greater clarity on what is needed to mitigate AI-enabled harms (Leslie et al., 2024).

## Conclusions

We conclude that OpenAI's April 2025 Preparedness Framework does not guarantee any AI risk mitigation practices. This conclusion is underpinned by three key findings: (1) the current Preparedness Framework ***requests***–without ***demanding***–evaluation of a small minority of AI risks, (2) the policy document ***encourages*** deployment of 'Medium' capabilities for severe harm, and (3) the policy document ***allows*** OpenAI's CEO to deploy even more dangerous capabilities, especially if other actors do so. For OpenAI, these results clarify concrete opportunities to more robustly mitigate harms enabled by OpenAI products. For the public, these results suggest that relying on self-regulation of AI developers will not reliably mitigate the harms of advanced AI systems. For AI safety research, results serve as a proof-of-concept for using affordance theory to clarify what policy documents and safety frameworks actually permit versus what they claim. Expanding such analyses for other AI interventions would support efforts to realise social benefits and mitigate social harms of advanced AI systems.

## Acknowledgements

We thank Dr. Dominic Glover, Dr. Marc Rigter, Professor Seth Lazar, and Dr. Sarah Bentley for their thoughtful and helpful contributions to this work. We also report that there are no funding sources or other potential competing interests to declare. We used Grammarly and Claude only to improve the wording of a minority of sentences included in this paper.

## Endnotes

[1] Note on language: Numerous AI developers and external policy documents use the term 'Frontier AI'. We recognise this is a contested term that has been critiqued for its murky definition, hype promotion, and potential colonial implications (Helfrich, 2024).

[2] The Preparedness Framework defines 'severe harm' as "the death or grave injury of thousands of people or hundreds of billions of dollars of economic damage" (OpenAI, 2025a).

[3] The 'MIT AI Risk Repository' (Slattery et al., 2025) has been cited 97 times according to 'Google Scholar' as of 20 September 2025.

[4] "Today, the OpenAI Board formed a Safety and Security Committee led by directors Bret Taylor (Chair), Adam D'Angelo, Nicole Seligman, and Sam Altman (CEO)" (OpenAI, 2024a).

[5] Defined by the MIT AI Risk Repository as "using AI systems to develop cyber weapons (e.g., coding cheaper, more effective malware), develop new or enhance existing weapons (e.g.,

Lethal Autonomous Weapons or CBRNE), or use weapons to cause mass harm" (Slattery et al., 2025). The Preparedness Framework ***requests*** evaluation of capabilities regarding "Cybersecurity" as well as Chemical weapons and biological weapons (OpenAI, 2025a). However, the Preparedness Framework only ***requests*** research into capabilities for developing nuclear and radiological weapons. Moreover, the Preparedness Framework does not ***request*** evaluation or research of capabilities for lethal autonomous weapons or high-yield explosives.

[6] Defined by the MIT AI Risk Repository as "AI developers or state-like actors competing in an AI 'race' by rapidly developing, deploying, and applying AI systems to maximize strategic or economic advantage, increasing the risk they release unsafe and error-prone systems" (Slattery et al., 2025). The Preparedness Framework ***encourages*** increased tolerance of deploying dangerous capabilities if another actor "develop[s] or release[s] a system with High or Critical capability in one of this Framework's Tracked Categories" (OpenAI, 2025a), as long as specific criteria are met[9].

[7] Defined by the MIT AI Risk Repository as "inadequate regulatory frameworks and oversight mechanisms failing to keep pace with AI development, leading to ineffective governance and the inability to manage AI risks appropriately" (Slattery et al., 2025). The Preparedness Framework ***requests*** evaluation of capabilities that "increase the rate at which new capabilities and risks emerge, to the point where our current oversight practices are insufficient to identify and mitigate new risks, including risks to maintaining human control of the AI system itself" (OpenAI, 2025a).

[8] Defined by the MIT AI Risk Repository as "AI systems that develop, access, or are provided with capabilities that increase their potential to cause mass harm through deception, weapons development and acquisition, persuasion and manipulation, political strategy, cyber-offense, AI development, situational awareness, and self-proliferation. These capabilities may cause mass harm due to malicious human actors, misaligned AI systems, or failure in the AI system" (Slattery et al., 2025). The Preparedness Framework only ***requests*** evaluation of AI capabilities for "AI Self-improvement", i.e., AI development.

[9] "We recognize that another frontier AI model developer might develop or release a system with High or Critical capability in one of this Framework's Tracked Categories… If we are able to rigorously confirm that such a scenario has occurred, then we could adjust accordingly the level of safeguards that we require in that capability area, but only if: • we assess that doing so does not meaningfully increase the overall risk of severe harm, • we publicly acknowledge that we are making the adjustment, • and, in order to avoid a race to the bottom on safety, we keep our safeguards at a level more protective than the other AI developer, and share information to validate this claim" (OpenAI, 2025a)

**References**

1. ABC News, 2025. OpenAI's ChatGPT to implement parental controls after teen's suicide [WWW Document] URL

https://www.abc.net.au/news/2025-09-03/chatgpt-to-implement-parental-controls-after-teen-suicide/105727518 (accessed 20 September 2025).
2. Bales, A., 2025. A polycrisis threat model for AI. *AI & SOCIETY*, pp.1-13.
3. Bengio, Y., Mindermann, S., Privitera, D., Besiroglu, T., Bommasani, R., Casper, S., Choi, Y., Fox, P., Garfinkel, B., Goldfarb, D., Heidari, H., Ho, A., Kapoor, S., Khalatbari, L., Longpre, S., Manning, S., Mavroudis, V., Mazeika, M., Michael, J., Newman, J., Ng, K.Y., Okolo, C.T., Raji, D., Sastry, G., Seger, E., Skeadas, T., South, T., Strubell, E., Tramèr, F., Velasco, L., Wheeler, N., Acemoglu, D., Adekanmbi, O., Dalrymple, D., Dietterich, T.G., Felten, E.W., Fung, P., Gourinchas, P.-O., Heintz, F., Hinton, G., Jennings, N., Krause, A., Leavy, S., Liang, P., Ludermir, T., Marda, V., Margetts, H., McDermid, J., Munga, J., Narayanan, A., Nelson, A., Neppel, C., Oh, A., Ramchurn, G., Russell, S., Schaake, M., Schölkopf, B., Song, D., Soto, A., Tiedrich, L., Varoquaux, G., Yao, A., Zhang, Y.-Q., Albalawi, F., Alserkal, M., Ajala, O., Avrin, G., Busch, C., Carvalho, A.C.P. de L.F. de, Fox, B., Gill, A.S., Hatip, A.H., Heikkilä, J., Jolly, G., Katzir, Z., Kitano, H., Krüger, A., Johnson, C., Khan, S.M., Lee, K.M., Ligot, D.V., Molchanovskyi, O., Monti, A., Mwamanzi, N., Nemer, M., Oliver, N., Portillo, J.R.L., Ravindran, B., Rivera, R.P., Riza, H., Rugege, C., Seoighe, C., Sheehan, J., Sheikh, H., Wong, D., Zeng, Y., 2025. International AI Safety Report. https://doi.org/10.48550/arXiv.2501.17805
4. Blue, G. and Hogan, M., 2024. Getting democracy wrong: How lessons from biotechnology can illuminate limits of the Asilomar AI principles. *Journal of Digital Social Research*, *6*(4), pp.107-117.
5. Brundage, M., Avin, S., Wang, J., Belfield, H., Krueger, G., Hadfield, G., Khlaaf, H., Yang, J., Toner, H., Fong, R., Maharaj, T., Koh, P.W., Hooker, S., Leung, J., Trask, A., Bluemke, E., Lebensold, J., O'Keefe, C., Koren, M., Ryffel, T., Rubinovitz, J.B., Besiroglu, T., Carugati, F., Clark, J., Eckersley, P., Haas, S. de, Johnson, M., Laurie, B., Ingerman, A., Krawczuk, I., Askell, A., Cammarota, R., Lohn, A., Krueger, D., Stix, C., Henderson, P., Graham, L., Prunkl, C., Martin, B., Seger, E., Zilberman, N., hÉigeartaigh, S.Ó., Kroeger, F., Sastry, G., Kagan, R., Weller, A., Tse, B., Barnes, E., Dafoe, A., Scharre, P., Herbert-Voss, A., Rasser, M., Sodhani, S., Flynn, C., Gilbert, T.K., Dyer, L., Khan, S., Bengio, Y., Anderljung, M., 2020. Toward Trustworthy AI Development: Mechanisms for Supporting Verifiable Claims. https://doi.org/10.48550/arXiv.2004.07213
6. Buhl, M.D., Bucknall, B., Masterson, T., 2025. Emerging Practices in Frontier AI Safety Frameworks. https://doi.org/10.48550/arXiv.2503.04746
7. Chang, C.C. and Lu, Y., 2025. Balancing Mission and Market: OpenAI's Struggle with Profit vs. Purpose. *Corp. & Bus. LJ*, *6*, p.1.
8. Crawford, K. and Joler, V., 2018. Anatomy of an AI System: The Amazon Echo as an anatomical map of human labor, data and planetary resources.Davis 2020
9. Davis, J.L., 2023. 'Affordances' for Machine Learning, in: Proceedings of the 2023 ACM Conference on Fairness, Accountability, and Transparency, FAccT '23. Association for Computing Machinery, New York, NY, USA, pp. 324–332. https://doi.org/10.1145/3593013.3594000
10. Davis, J.L., 2020. *How artifacts afford: The power and politics of everyday things*. MIT Press.

11. Davis, J.L., Chouinard, J.B., 2016. Theorizing Affordances: From Request to Refuse. *Bulletin of Science, Technology & Society 36*, 241–248. https://doi.org/10.1177/0270467617714944
12. Digital Transformation Agency, 2025. Evaluation of the whole-of-government trial of Microsoft 365 Copilot: full report [WWW Document] URL https://www.digital.gov.au/initiatives/copilot-trial/microsoft-365-copilot-evaluation-report-full (accessed 20 September 2025).
13. DSIT, 2025. Frontier AI Safety Commitments, AI Seoul Summit 2024 [WWW Document] URL https://www.gov.uk/government/publications/frontier-ai-safety-commitments-ai-seoul-summit-2024/frontier-ai-safety-commitments-ai-seoul-summit-2024 (accessed 20 September 2025).
14. EOP, 2025. America's AI Action Plan https://www.whitehouse.gov/wp-content/uploads/2025/07/Americas-AI-Action-Plan.pdf (accessed 20 September 2025).
15. EU, 2024. Regulation (EU) 2024/1689 of the European Parliament and of the Council of 13 June 2024 laying down harmonised rules on artificial intelligence and amending Regulations (EC) No 300/2008, (EU) No 167/2013, (EU) No 168/2013, (EU) 2018/858, (EU) 2018/1139 and (EU) 2019/2144 and Directives 2014/90/EU, (EU) 2016/797 and (EU) 2020/1828 (Artificial Intelligence Act) (Text with EEA relevance). https://eur-lex.europa.eu/eli/reg/2024/1689/oj (Accessed 20 September 2025).
16. Fazelpour, S., Magnani, M., 2025. Aspirational Affordances of AI. https://doi.org/10.48550/arXiv.2504.15469
17. Frontier Model Forum, 2025. Publications - Public Comments [WWW Document] URL https://www.frontiermodelforum.org/publications/#public-comments (Accessed 29 September 2025).
18. Gibson, J. J., 1979. *The ecological approach to visual perception*. Boston, MA: Houghton Mifflin.
19. Glover, D., 2022. Affordances and agricultural technology. *Journal of Rural Studies* 94, 73–82. https://doi.org/10.1016/j.jrurstud.2022.05.007
20. Goodyear, P., 2022. Realising the good university: Social innovation, care, design justice and educational infrastructure. *Postdigital Science and Education*, *4*(1), pp.33-56.
21. Grey, M., Segerie, C.-R., 2025. Safety by Measurement: A Systematic Literature Review of AI Safety Evaluation Methods. https://doi.org/10.48550/arXiv.2505.05541
22. Helfrich, G., 2024. The harms of terminology: why we should reject so-called "frontier AI". *AI and Ethics*, *4*(3), pp.699-705.
23. Henman, P.W.F., 2022. Digital social policy: Past, present, future. *Journal of Social Policy*, *51*(3), pp.535-550.
24. Kapoor, S., Kolt, N., Lazar, S., 2025. Build Agent Advocates, Not Platform Agents. https://doi.org/10.48550/arXiv.2505.04345
25. Kasirzadeh, A., 2025. Two types of AI existential risk: decisive and accumulative. *Philosophical Studies*, 182, pp.1975–2003. https://doi.org/10.1007/s11098-025-02301-3

26. Kiviat, B., 2023. The moral affordances of construing people as cases: How algorithms and the data they depend on obscure narrative and noncomparative justice. *Sociological Theory*, *41*(3), pp.175-200.
27. Leslie, D., Ashurst, C., González, N.M., Griffiths, F., Jayadeva, S., Jorgensen, M., Katell, M., Krishna, S., Kwiatkowski, D., Martins, C.I. and Mahomed, S., 2024. 'Frontier AI, 'Power, and the Public Interest: Who benefits, who decides?. *Harvard Data Science Review*, (Special Issue 5).
28. Norman, D. 2013. *The Design of Everyday Things*. Basic Books, New York.
29. OpenAI, 2025a. Our updated Preparedness Framework [WWW Document] URL https://openai.com/index/updating-our-preparedness-framework/ (accessed 20 September 2025).
30. OpenAI, 2025b. Our structure [WWW Document] URL https://openai.com/our-structure/ (accessed 20 September 2025).
31. OpenAI, 2025c. OpenAI o3 and o4-mini System Card [WWW Document] URL https://openai.com/index/o3-o4-mini-system-card/ (accessed 20 September 2025).
32. OpenAI, 2024a. OpenAI board forms safety and security committee [WWW Document] URL https://openai.com/index/openai-board-forms-safety-and-security-committee/ (accessed 20 September 2025).
33. OpenAI, 2024b. OpenAI o1 System Card [WWW Document] URL https://openai.com/index/openai-o1-system-card/ (accessed 20 September 2025).
34. OpenAI, 2023. Preparedness Framework (Beta). https://www.google.com/url?sa=t&source=web&rct=j&opi=89978449&url=https://cdn.openai.com/openai-preparedness-framework-beta.pdf&ved=2ahUKEwjhoP3SiMiOAxXBrVYBHcfCHaUQFnoECBcQAQ&usg=AOvVaw2LNyr7jICjP2bqjGaqVIpv (accessed 20 September 2025).
35. OpenClipart, 2025. Vector drawing of pdf file type computer icon [WWW Image] URL https://freesvg.org/vector-drawing-of-pdf-file-type-computer-icon (accessed 29 September 2025).
36. Pistillo, M., 2025. Towards Frontier Safety Policies Plus. https://doi.org/10.48550/arXiv.2501.16500
37. Raman, D., Madkour, N., Murphy, E.R., Jackson, K., Newman, J., 2025. Intolerable Risk Threshold Recommendations for Artificial Intelligence. https://doi.org/10.48550/arXiv.2503.05812
38. Ruster, L.P., Davis, J.L., 2025. The Gaps that Never Were: Reconsidering Responsible AI's Principle-Practice Problem, in: Proceedings of the 2025 ACM Conference on Fairness, Accountability, and Transparency, FAccT '25. Association for Computing Machinery, New York, NY, USA, pp. 350–360. https://doi.org/10.1145/3715275.3732024
39. Slattery, P., Saeri, A.K., Grundy, E.A.C., Graham, J., Noetel, M., Uuk, R., Dao, J., Pour, S., Casper, S., Thompson, N., 2025. The AI Risk Repository: A Comprehensive Meta-Review, Database, and Taxonomy of Risks From Artificial Intelligence. https://arxiv.org/abs/2408.12622v2
40. Stelling, L., Yang, M., Gipiškis, R., Staufer, L., Chin, Z.S., Campos, S., Chen, M., 2025. Existing Industry Practice for the EU AI Act's General-Purpose AI Code of Practice Safety and Security Measures. https://doi.org/10.48550/arXiv.2504.15181

41. Swant, 2025. A roundup of AI psychosis stories [WWW Document] URL https://www.transformernews.ai/p/ai-psychosis-stories-roundup (accessed 20 September 2025).
42. Zech, P., Haller, S., Lakani, S.R., Ridge, B., Ugur, E., Piater, J., 2017. Computational models of affordance in robotics: a taxonomy and systematic classification. *Adaptive Behavior* 25, 235–271. https://doi.org/10.1177/1059712317726357